\documentclass[12pt]{iopart}
\usepackage{iopams} 
\usepackage{bbm, bm, mathtools, hyperref, graphicx, xcolor, float, biblatex}
\begin{document}

\title[Integrable Fractional mKdV, sineG, and sinhG Equations]{Integrable Fractional Modified Korteweg-de Vries, Sine-Gordon, and Sinh-Gordon Equations}

\author{Mark J. Ablowitz}
\address{Department of Applied Mathematics, University of Colorado, Boulder, Colorado, U.S.A.}
\ead{mark.ablowitz@colorado.edu}

\author{Joel B. Been}
\address{Department of Applied Mathematics and Statistics and Department of Physics, Colorado School of Mines, Golden, Colorado, U.S.A.}
\ead{joelbeen@mines.edu}

\author{Lincoln D. Carr}
\address{Quantum Engineering Program, Department of Applied Mathematics and Statistics, and Department of Physics, Colorado School of Mines, Golden, Colorado, U.S.A.}
\ead{lcarr@mines.edu}

\vspace{10pt}
\begin{indented}
\item[]March 2022
\end{indented}

\begin{abstract}
    The inverse scattering transform allows explicit construction of solutions to many physically significant nonlinear wave equations. Notably, this method can be extended to fractional nonlinear evolution equations characterized by anomalous dispersion using completeness of suitable eigenfunctions of the associated linear scattering problem. In anomalous diffusion, the mean squared displacement is proportional to $t^{\alpha}$, $\alpha>0$, while in anomalous dispersion, the speed of localized waves is proportional to $A^{\alpha}$, where $A$ is the amplitude of the wave. Fractional extensions of the modified Korteweg-deVries (mKdV), sine-Gordon (sineG) and sinh-Gordon (sinhG) and associated hierarchies are obtained.  Using  symmetries present in the  linear scattering problem,  these equations can be connected with a scalar family of nonlinear evolution equations of which fractional mKdV (fmKdV), fractional sineG (fsineG), and fractional sinhG (fsinhG) are special cases. Completeness of solutions to the scalar problem is obtained and, from this, the nonlinear evolution equation is characterized in terms of a spectral expansion. In particular, fmKdV, fsineG, and fsinhG are explicitly written. One-soliton solutions are derived for fmKdV and fsineG using the inverse scattering transform and these solitons are shown to exhibit anomalous dispersion.

\end{abstract}

\maketitle

\section{Introduction}

Fractional calculus has been effectively applied to describe physical systems with anomalous behavior associated with multi-scale media. The underlying fractional mathematical formulation, originally designed to interpolate between integer derivative orders, has been used to describe new phenomena, such as novel forms of transport in biology \cite{saxton_2007,bronstein_2009,weigel_2011,regner_2013}, amorphous materials \cite{scher_1975,pfister_1977,gu_1996}, porous media \cite{benson_2000,benson_2001,meerschaert_2008}, and climate science \cite{koscielny_1998} amongst others. Fractional equations often predict physically measurable quantities that follow power laws. For example, in anomalous diffusion, the mean squared displacement is related to time by a power law $t^{\alpha}$, $\alpha > 0$ \cite{west_1987,random_walk,west_1997,anomalous_issue}. Similarly, for integrable soliton equations, fractional generalizations predict \emph{anomalous dispersion} where the speed of localized solitonic waves are related to their amplitude by a power law \cite{fKdV_fNLS}.

Integrable evolution equations are key elements in the study of nonlinear dynamics because they have deep mathematical structure and provide exactly solvable models whose results can be compared with experimental and numerical data. Some important and well-known examples of integrable evolution equations are the Korteweg-de Vries (KdV), modified Korteweg-de Vries (mKdV), nonlinear Schrodinger (NLS), sine-Gordon (sineG), and sinh-Gordon (sinhG) equations. These equations are solvable by the inverse scattering transformation (IST), a nonlinear generalization of Fourier transforms where the nonlinear equation is associated with a linear scattering problem. They also admit an infinite set of conservation laws and have soliton solutions which are robust localized  traveling waves \cite{Ablowitz1,Ablowitz3}.

In \cite{fKdV_fNLS}, we obtained and analyzed the integrable fractional Korteweg-de Vries (fKdV) and integrable fractional nonlinear Schr\"odinger (fNLS) equations. These were two examples of a hierarchy of fractional equations that can be constructed. In the case of NLS, the hierarchy is written in terms of $2 \times 2$ matrix operators. In this article, we demonstrate that this process can be applied to define and solve key, physically relevant nonlinear evolution equations --- namely the fmKdV, fsineG, and fsinhG equations in terms of scalar operators. Although the fmKdV, fsineG, and fsinhG equations can be written in terms of matrix operators the scalar system is considerably simpler, more compact and, provides a direct analog of the scalar fKdV operator.

The fractional operators of these integrable systems are are nonlinear generalizations of the well-established Riesz fractional derivative. Although there are many fractional derivatives, the Riesz formulation is particularly intuitive and accessible for physicists who do not specialize in this area of mathematics.  The Riesz fractional derivative is defined by its Fourier multiplier $|k|^{2 \epsilon}$, $0<\epsilon < 1$, and can be understood as the fractional power of $- \partial_{x}^2$. Fractional equations defined using the Riesz fractional derivative (alternately termed the Riesz transform \cite{riesz} or fractional Laplacian \cite{what_is_frac_lap}) are effective tools when describing behavior in complex systems because the Riesz fractional derivative is closely related to non-Gaussian statistics \cite{frac_stoch}. It has found physical applications in describing movement of water in porous media \cite{frac_porous}, transport of temperature in fluid dynamics \cite{quasi_geo}, and power law attenuation in materials \cite{power_law_attn} amongst many others \cite{frac_lap,diff_appl_1,diff_appl_2}.

The KdV equation describes quadratic nonlinear waves with weak dispersion; it was discovered in water waves over one hundred years ago \cite{KDV}. The KdV equation admits solitary wave solutions which are localized waves of permanent form that propagate unidirectionally and whose speed and amplitude are linearly related. Seventy years later, using numerical methods, KdV solitary waves were found to interact elastically; they were termed solitons \cite{zabusky1965interaction}. Soon afterwards the KdV equation with decaying initial data was linearized and soliton solutions were obtained analytically using inverse scattering methods \cite{gardner1967method}. A few years later the NLS equation was found to be solvable via inverse scattering and to have soliton solutions. In \cite{AKNS}, the linearization procedure was generalized with the NLS, mKdV, and sineG (in light cone coordinates) equations as special cases. The procedure was termed the Inverse Scattering Transform (IST). These equations arise in numerous physical contexts \cite{Ablowitz2,Ablowitz2011}.  Remarkably, all of these equations have fractional extensions which pave the way for applications to anomalous dispersion and multi-scale behavior.

In this article, we define and solve the fmKdV, fsineG, and fsinhG equations on the line with suitable initial data using three ingredients: a general nonlinear equation solvable by the IST, a completeness relation for squared eigenfunctions, and an anomalous dispersion relation. We develop a scalar reduction of the Ablowitz-Kaup-Newell-Segur (AKNS) system in which we find the fmKdV, fsineG, and fsinhG equations as special cases using power law dispersion relations. Then, we characterize a completeness relation for squared eigenfunctions of this scalar system. This completeness relation provides a spectral representation for the fractional operators in the fmKdV, fsineG, and fsinhG equations, giving the equations an explicit representation in physical space. From basic IST theory we derive a one-soliton solutions to the whole class of nonlinear equations described by the scalar reduction; in particular, we give those for the fmKdV and fsineG equations. 
We also verify these results using the completeness relation for squared eigenfunctions. The velocity of these solitons are related to their amplitude by a power law. Therefore, the fmKdV and fsineG equations predict anomalous dispersion.

\section{AKNS Scattering System and Scalar Reduction}

The inverse scattering transformation relies on associating the nonlinear problem we want to solve to a linear scattering problem by taking the potential of the linear problem to be the solution of the nonlinear problem. For many nonlinear evolution equations, e.g., the mKdV, sineG, and NLS equations, the associated linear scattering problem is the AKNS system (also often called the AKNS eigenvalue problem). The nonlinear evolution equations are linearized by the scattering problem. We previously demonstrated that the AKNS system can linearize the \emph{fractional} Nonlinear Schr\"odinger equation \cite{fKdV_fNLS} via a $2 \times 2$ matrix operator. \\

Here, we will show that given a symmetry reduction, the vector valued nonlinear evolution equations associated to the AKNS scattering problem becomes a scalar nonlinear evolution equation. This family of equations is then shown to contain the mKdV, sineG, and sinhG equations:
\begin{align}
    q_{t} &\mp 6 q^2 q_{x} + q_{xxx} = 0, \label{eqn:KdV} \\
    u_{xt} &= \sin{u}, \label{eqn:sine_gordon}
\end{align}
where $r = \pm q$ for KdV and $r = -q$, with $u_{x}/2 = -q$ for sineG with $q$ real. We also note that with $r = q=u_x/2$ and $q$ real we find the sinhG equation: 
\begin{equation}
u_{xt}=\sinh u
 \label{eqn:sinh_gordon}
\end{equation}
Below, we show that this family of scalar evolution equations also contains fmKdV, fsineG, and fsinhG as well as their hierarchies. First, we will outline scattering theory of the AKNS system and show how this leads to the scalar scattering problem.

\subsection{AKNS Scattering Problem}

The Ablowitz-Kaup-Newell-Segur (AKNS) system is the $2\times2$ scattering problem for the vector-valued function $\mathbf{v}(x) = \left(v_{1}(x),v_{2}(x)\right)^{T}$ ($T$ represents transpose)
\begin{align}
    v^{(1)}_{x} &= - i k v^{(1)} + q(x,t) v^{(2)}, \label{eqn:AKNS_scattering_1} \\
    v^{(2)}_{x} &= +i k v^{(2)} + r(x,t) v^{(1)}, \label{eqn:AKNS_scattering_2}
\end{align}
where $q$ and $r$ act as potentials and $k$ is an eigenvalue. We can associate to this scattering problem a vector-valued family of integrable nonlinear equations \cite{AKNS}
\begin{align}
    \sigma_{3}\partial_t\mathbf{u} + 2 A_{0}(\mathbf{L}^{A}) \mathbf{u} = 0, \quad \sigma_{3} = \begin{pmatrix} 1 & 0 \\ 0 & -1 \end{pmatrix},
    \label{eqn:AKNS_general_evolution_equation}
\end{align}
where $\mathbf{u} = \left( r, q \right)^{T}$ decays sufficiently rapidly at infinity and the $2\times 2$ matrix operator 
\begin{align}
    \mathbf{L}^{A} \equiv \frac{1}{2 i} \begin{pmatrix} \partial_{x} - 2 r I_{-} q & 2 r I_{-} r \\
    - 2 q I_{-} q & -\partial_{x} + 2 q I_{-} r \end{pmatrix}, \label{eqn:AKNS_L_operator_adjoint} 
\end{align}
with $I_{-} = \int_{-\infty}^{x} dy$. $\mathbf{L}^{A}$ is the adjoint of
\begin{align}
    \mathbf{L} \equiv \frac{1}{2 i} \begin{pmatrix} -\partial_{x} - 2 q I_{+} r & -2 q I_{+} q \\
    2 r I_{+} r & \partial_{x} + 2 r I_{+} q \end{pmatrix}, \label{eqn:AKNS_L_operator}
\end{align}
with $I_{+} = \int_{x}^{\infty} dy$. The function $A_{0}$ has been traditionally considered to be meromorphic. The family of equations represented by (\ref{eqn:AKNS_general_evolution_equation}) is commonly related to cases when $A_{0}(\mathbf{L}^{A})=\left(\mathbf{L}^{A}\right)^n$, $n=1,2,3...$. However, using the completeness relation for squared eigenfunctions which is discussed in the next section, it was shown that this can be extended to much more general $A_{0}(k)$ \cite{fKdV_fNLS}. The operator $A_{0}(\mathbf{L}^{A})$ can also be related to the dispersion relation of the linearization of (\ref{eqn:AKNS_general_evolution_equation}). Specifically, if we put 
$q = e^{i (k x - \omega((k) t)}$  into the linearization of (\ref{eqn:AKNS_general_evolution_equation}), we have
\begin{align}
    A_{0}\left(\frac{k}{2}\right) = -\frac{i}{2} \omega(-k). \label{eqn:AKNS_dispersion_relation}
\end{align}
We can obtain the Nonlinear Schr\"odinger equation from equation (\ref{eqn:AKNS_general_evolution_equation}) by putting $r = \mp q^{*}$ with its linear dispersion relation $\omega(k) = - k^2$. Similarly, sineG, sinhG, and mKdV follow from $r = - q$, 
$\omega(k) = k^{-1}$; $r = + q$, 
$\omega(k) = k^{-1}$; and $r = \pm q$, $\omega(k) = - k^3$, respectively with $q$ real. In \cite{fKdV_fNLS}, it was shown that fNLS could be obtained from $r = \pm q$, $\omega(k) = - k^2 |k|^{\epsilon}$. The associated hierarchy of integrable equations follows by taking $\omega(k) = - k^n |k|^{\epsilon}, n=3,4,...$.

We will take the linearization of fmKdV, fsineG, and fsinhG to be
\begin{align}
    q_{t} + (- \partial_{x}^2)^{\epsilon} q_{xxx} &= 0, \label{eqn:linear_mKdV} \\
    u_{t x} = (- \partial_{x}^2)^{\epsilon} u, ~~ &q_{t} = \int_{-\infty}^{x} (- \partial_{\xi}^2)^{\epsilon}q(\xi,t)\, d\xi, \label{eqn:linear_SG}
\end{align}
where $(- \partial_{x}^2)^{\epsilon}$ is the Riesz fractional derivative. 
Notice that both fsineG and fsinhG have the same linear equation (\ref{eqn:linear_SG}). The linearization of fKdV has dispersion relation $\omega(k) = - k^3 |k|^{2 \epsilon}$ and that of fsineG and fsinhG is $\omega(k) = |k|^{2\epsilon}/k$. Therefore, fKdV can be obtained from (\ref{eqn:AKNS_general_evolution_equation}) with $r = \pm q$ and $A_{0}(k) = - 4 i k^3 |2 k|^{2\epsilon}$ and similarly fsineG (sinhG) are (\ref{eqn:AKNS_general_evolution_equation}) with $r = -q$ ($r = +q$) and $A_{0}(k) = i |2 k|^{2\epsilon}/(4 k)$. 

\subsection{Scattering Data for the AKNS System}

With sufficient decay and smoothness of $\mathbf{u}$, we define eigenfunctions functions for the AKNS system as solutions to equations (\ref{eqn:AKNS_scattering_1}) and (\ref{eqn:AKNS_scattering_2}) satisfying the boundary conditions
\begin{alignat}{3}
    \boldsymbol{\phi}(x;k,t) &\sim \begin{pmatrix}
    1 \\
    0
    \end{pmatrix} e^{-i k x},~~\overline{\boldsymbol{\phi}}(x;k,t) &&\sim \begin{pmatrix}
    0 \\
    1
    \end{pmatrix} e^{+i k x}, ~~ &&x\to-\infty, \label{eqn:AKNS_asymptotic_phi} \\
    \boldsymbol{\psi}(x;k,t) &\sim \begin{pmatrix}
    0 \\
    1
    \end{pmatrix} e^{+i k x},~~\overline{\boldsymbol{\psi}}(x;k,t) &&\sim \begin{pmatrix}
    1 \\
    0
    \end{pmatrix} e^{-i k x}, ~~ &&x\to+\infty. \label{eqn:AKNS_asymptotic_psi}
\end{alignat}
As the eigenfunctions $\boldsymbol{\psi}$, $\overline{\boldsymbol{\psi}}$ are linearly independent, we can write $\boldsymbol{\phi}$ and $\overline{\boldsymbol{\phi}}$ as
\begin{align}
    \boldsymbol{\phi}(x;k,t) &= b(k,t) \boldsymbol{\psi}(x;k,t) + a(k,t) \overline{\boldsymbol{\psi}}(x;k,t), \label{eqn:AKNS_scattering_data_phi} \\
    \overline{\boldsymbol{\phi}}(x;k,t) &= \overline{a}(k,t) \boldsymbol{\psi}(x;k,t) + \overline{b}(k,t) \overline{\boldsymbol{\psi}}(x;k,t). \label{eqn:AKNS_scattering_data_psi}
\end{align}
Then, we can write the scattering data explicitly in terms of the eigenfunctions as
\begin{alignat}{2}
    a(k,t) &= W(\boldsymbol{\phi}, \boldsymbol{\psi}), \quad \overline{a}(k,t) &&= W(\overline{\boldsymbol{\psi}},\overline{\boldsymbol{\phi}}), \label{eqn:AKNS_wronskian_relations_a} \\
    b(k,t) &= W(\overline{\boldsymbol{\psi}},\boldsymbol{\phi}), \quad \overline{b}(k,t) &&= W(\overline{\boldsymbol{\phi}},\boldsymbol{\psi}), \label{eqn:AKNS_wronskian_relations_b}
\end{alignat}
with the Wronskian given by $W(u,v) = u^{(1)} v^{(2)} - u^{(2)}v^{(1)}$. The transmission and reflection coefficients, $\tau(k,t)$, $\overline{\tau}(k,t)$ and $\rho(k,t)$, $\overline{\rho}(k,t)$, are defined by
\begin{alignat}{2}
    \tau(k,t) &= \frac{1}{a(k,t)}, \quad &&\rho(k,t) = \frac{b(k,t)}{a(k,t)}, \label{eqn:AKNS_tau_rho}  \\
    \overline{\tau}(k,t) &= \frac{1}{\overline{a}(k,t)}, \quad &&\overline{\rho}(k,t) = \frac{\overline{b}(k,t)}{\overline{a}(k,t)}. \label{eqn:AKNS_tau_rho_bar}
\end{alignat}
We also define the mixed reflection coefficient by
\begin{align}
    \tilde{\rho}(k,t) &= \frac{\overline{b}(k,t)}{a(k,t)}. \label{eqn:AKNS_rho_tilde}
\end{align}
The zeros of $a$ and $\overline{a}$ at $k_{j} = \xi_{j} + i \eta_{j}$, $\eta_{j} > 0$, $j = 1, 2, ..., J$ and $\overline{k}_{j} = \overline{\xi}_{j} + i \overline{\eta}_{j}$, $\overline{\eta}_{j} < 0$, $j = 1, 2, ..., \overline{J}$, respectively, are eigenvalues of the AKNS system corresponding to bound states. With decaying data, these eigenvalues exist only when $r=-q$. We assume the eigenvalues are `proper', i.e., they are simple zeros of $a$/$\overline{a}$, they are not on the real $k$ axis, and $J= \overline{J}$; cf. \cite{Ablowitz3}. The bound state eigenfunctions are related by
\begin{align}
    \boldsymbol{\phi}_{j}(x,t) = b_{j}(t) \boldsymbol{\psi}_{j}(x,t), \quad \overline{\boldsymbol{\phi}}_{j}(x,t) = \overline{b}_{j}(t) \overline{\boldsymbol{\psi}}_{j}(x,t),
\end{align}
where $b_{j}(t) = b(k_{j},t)$. We also define the norming constants by
\begin{align}
    C_{j}(t) &= b_{j}(t)/a_{j}'(t), \quad \overline{C}_{j}(t) = \overline{b}_{j}(t)/\overline{a}_{j}'(t), \label{eqn:AKNS_norming_constants} \\
    \tilde{C}_{j}(t) &= \overline{b}_{j}(t)/a_{j}'(t), \label{eqn:AKNS_norming_constants_tilde}
\end{align}
where $a_{j}'(t) = \partial_{k} a(k,t) |_{k = k_{j}}$, etc.
When $r = \mp q^{*}$ in (\ref{eqn:AKNS_scattering_1})-(\ref{eqn:AKNS_scattering_2}), we have the symmetry reductions
\begin{align}
    \overline{\boldsymbol{\psi}}(x,k,t) &= \sigma \boldsymbol{\psi}^{*}(x,k^{*},t), \quad \overline{\boldsymbol{\phi}}(x,k,t) = \sigma^{-1} \boldsymbol{\phi}^{*}(x,k^{*},t),
\end{align}
for the eigenfunctions and 
$\overline{a}(k,t) = a^{*}(k^{*},t)$ and $\overline{b}(k,t) = \mp b^{*}(k^{*},t)$ for the scattering data where
\begin{align}
    \sigma_{\pm} = \begin{pmatrix}0&1\\\pm1&0\end{pmatrix}, \quad \sigma_{\pm}^{-1} = \begin{pmatrix}0&\pm1\\\ 1&0\end{pmatrix}.
\end{align}
When $r = \pm q$, $q$ real, we have the symmetry reductions
\begin{alignat}{2}
    \overline{\boldsymbol{\psi}}(x;k,t) &= \sigma_{\pm} \boldsymbol{\psi}(x;-k,t), \quad &&\overline{\boldsymbol{\phi}}(x;k,t) = \sigma_{\pm}^{-1} \boldsymbol{\phi}(x;-k,t), \label{eqn:scalar_symmetry_reduction_eigenfunctions}
\end{alignat}
for the eigenfunctions and 
\begin{alignat}{2}
    \overline{a}(k,t) &= a(-k,t), \quad &&\overline{b}(k,t) = \pm b(-k,t), \label{eqn:scalar_symmetry_reduction_data}
\end{alignat}
for the scattering data. From the scattering eigenfunctions $\boldsymbol{\psi}$ and $\boldsymbol{\phi}$, we can construct the eigenfunctions of the operator $\mathbf{L}$, $\mathbf{\Psi}(x,k,t)$ and $\overline{\mathbf{\Psi}}(x,k,t)$, and its adjoint $\mathbf{L}^{A}$, $\mathbf{\Psi}^{A}(x,k,t)$ and $\overline{\mathbf{\Psi}}^{A}(x,k,t)$ by
\begin{alignat}{2}
    \mathbf{\Psi}(x,k,t) &= \left((\psi^{(1)}(x,k,t))^2, (\psi^{(2)}(x,k,t))^2 \right)^{T}, \label{eqn:bold_psi_definition} \\ ~ \overline{\mathbf{\Psi}}(x,k,t) &= \left((\overline{\psi}^{(1)}(x,k,t))^2, (\overline{\psi}^{(2)}(x,k,t))^2 \right)^{T}, \label{eqn:bold_psi_bar_definition}  \\
    \mathbf{\Psi}^{A}(x,k,t) &= \left((\phi^{(2)}(x,k,t))^2, -(\phi^{(1)}(x,k,t))^2 \right)^{T}, \label{eqn:bold_psi_A_definition} \\ \overline{\mathbf{\Psi}}^{A}(x,k,t) &= \left((\overline{\phi}^{(2)}(x,k,t))^2, -(\overline{\phi}^{(1)}(x,k,t))^2 \right)^{T}. \label{eqn:bold_psi_bar_A_definition}
\end{alignat}
Notice that these are all written in terms of squared eigenfunctions of the AKNS system. Explicitly, we have
\begin{alignat}{2}
    \mathbf{L} \mathbf{\Psi} &= k \mathbf{\Psi}, ~~ &&\mathbf{L} \overline{\mathbf{\Psi}} = k \overline{\mathbf{\Psi}}, \label{eqn:AKNS_eigenfunctions} \\
    \mathbf{L}^{A} \mathbf{\Psi}^{A} &= k \mathbf{\Psi}^{A}, ~~ &&\mathbf{L}^{A} \overline{\mathbf{\Psi}}^{A} = k \overline{\mathbf{\Psi}}^{A}. \label{eqn:AKNS_adjoint_eigenfunctions}
\end{alignat}

\subsection{Scalar Scattering System}

The scattering equations for the scalar system, obtained from the symmetry reduction $r = \pm q$, are
\begin{align}
    v^{(1)}_{x} &= - i k v^{(1)} + q(x,t) v^{(2)}, \label{eqn:scalar_scattering_1} \\
    v^{(2)}_{x} &= +i k v^{(2)} \pm q(x,t) v^{(1)}. \label{eqn:scalar_scattering_2}
\end{align}
To construct a family of nonlinear evolution equations for this system, which is a subset of the class in equation (\ref{eqn:AKNS_general_evolution_equation}), we use the eigenvalue relations in equations (\ref{eqn:AKNS_eigenfunctions}) and (\ref{eqn:AKNS_adjoint_eigenfunctions}) in addition to an orthogonality relation from the AKNS system. Taking $r = \pm q$ and writing out $\mathbf{L} \mathbf{\Psi} = k \mathbf{\Psi}$ in components, we have
\begin{align}
    2 i k ( \psi^{(1)})^2 &= -\frac{\partial ( \psi^{(1)})^2}{\partial x} - 2 q I_{+}\left[ q (\pm ( \psi^{(1)})^2 + ( \psi^{(2)})^2) \right], \label{eqn:p_plus_derivation_1} \\
    2 i k ( \psi^{(2)})^2 &= \frac{\partial ( \psi^{(2)})^2}{\partial x} + 2 q I_{+}\left[ q (( \psi^{(1)})^2 \pm ( \psi^{(2)})^2) \right]. \label{eqn:p_plus_derivation_2}
\end{align}
We define the functions
\begin{align}
    \mu_{-}(x,k) &= ( \psi^{(1)}(x,k))^2 + ( \psi^{(2)}(x,k))^{2}, \label{eqn:mu_plus} \\
    \mu_{+}(x,k) &= ( \psi^{(1)}(x,k))^2 - ( \psi^{(2)}(x,k))^{2}. \label{eqn:mu_minus}
\end{align}
Taking $r = +q$, adding (\ref{eqn:p_plus_derivation_1}) to (\ref{eqn:p_plus_derivation_2}) and using (\ref{eqn:mu_plus}) and (\ref{eqn:mu_minus}) we have
\begin{align}
    2 i k \mu_{-} = -\frac{\partial \mu_{+}}{\partial x}. \label{eqn:p_plus_derivation_3}
\end{align}
Subtracting equation (\ref{eqn:p_plus_derivation_2}) from (\ref{eqn:p_plus_derivation_1}) yields
\begin{align}
    2 i k \mu_{+} = -\frac{\partial \mu_{-}}{\partial x} - 4 q I_{+} \left[ q \mu_{-} \right]. \label{eqn:p_plus_derivation_4}
\end{align}
Putting equation (\ref{eqn:p_plus_derivation_3}) into (\ref{eqn:p_plus_derivation_4}) we get the following scalar eigenvalue equation
\begin{alignat}{2}
    L_{+} \mu_{+} &= k^2 \mu_{+}, ~~ &&L_{+} = - \frac{1}{4} \frac{\partial^2}{\partial x^2} + q^2 + q I_{+} q_{y}. \label{eqn:mu_eigenvalue_equation_plus}
\end{alignat}
We can similarly show that
\begin{alignat}{2}
    L_{+}^{A} \nu_{+} &= k^2 \nu_{+}, ~~ &&L_{+}^{A} = - \frac{1}{4} \frac{\partial^2}{\partial x^2} + q^2 + q_{x} I_{-}, \label{eqn:nu_eigenvalue_equation_plus}
\end{alignat}
where
\begin{align}
    \nu_{+}(x,k) &= (\phi^{(1)}(x,k))^{2} + (\phi^{(2)}(x,k))^{2}, \\
    \nu_{-}(x,k) &= (\phi^{(1)}(x,k))^{2} - (\phi^{(2)}(x,k))^{2}.
\end{align}
For $r = - q$, we have
\begin{alignat}{2}
    L_{-} \mu_{-} &= k^2 \mu_{-}, ~~ &&L_{-} = - \frac{1}{4} \frac{\partial^2}{\partial x^2} - q^2 - q I_{+} q_{y}, \label{eqn:mu_eigenvalue_equation_minus} \\
    L_{-}^{A} \nu_{-} &= k^2 \nu_{-}, ~~ &&L_{-}^{A} = - \frac{1}{4} \frac{\partial^2}{\partial x^2} - q^2 - q_{x} I_{-} q. \label{eqn:nu_eigenvalue_equation_minus}
\end{alignat}
Equations (\ref{eqn:mu_eigenvalue_equation_plus}), (\ref{eqn:nu_eigenvalue_equation_plus}), (\ref{eqn:mu_eigenvalue_equation_minus}), and (\ref{eqn:nu_eigenvalue_equation_minus}) define the operators and squared eigenfunctions of the scalar scattering system. For the AKNS scattering problem, we know that the following orthogonality relation holds \cite{AKNS}
\begin{align}
    \int_{-\infty}^{\infty} \left\{ \left( r_{t} + 2 \Omega(k) r \right) ( \psi^{(1)})^2 + \left( - q_{t} + 2 \Omega(k) q \right) ( \psi^{(2)})^{2} \right\} dx = 0,
\end{align}
where $\Omega(k)$ is a suitable function of $k$, taken to be meromorphic in  \cite{ablowitz1974inverse}. With $r = \pm q$, $\Omega(k) = i k \Theta(k^2)$, and $\mu_{1}$ and $\mu_{2}$ in (\ref{eqn:mu_plus}) and (\ref{eqn:mu_minus}), we may write
\begin{align}
    \int_{-\infty}^{\infty} \left\{ q_{t} \mu_{\pm} + 2 q i k \Theta(k^2) \mu_{\mp} \right\} dx = 0.
\end{align}
Noting that we have
\begin{align}
    2 i k \mu_{-} = - \frac{\partial \mu_{+}}{\partial x} \text{ for } r = +q, \\
    2 i k \mu_{+} = - \frac{\partial \mu_{-}}{\partial x} \text{ for } r = -q,
\end{align}
and using the extension of equations (\ref{eqn:mu_eigenvalue_equation_plus}) and (\ref{eqn:mu_eigenvalue_equation_minus})
\begin{alignat}{2}
    \Theta(L_{\pm}) \mu_{\pm} &= \Theta(k^2) \mu_{\pm}, \label{eqn:L_eigenvalue_equation}
\end{alignat}
we can write
\begin{align}
    \int_{-\infty}^{\infty} \left\{ q_{t} \mu_{\pm} + q_{x} \Theta(L_{\pm}) \mu_{\pm} \right\} dx = 0,
\end{align}
using integration by parts. We can then shift $\Theta(L_{\pm})$ from operating on $\mu_{\pm}$ to $q_x$ using the adjoint of $L$,
\begin{align}
    L_{\pm}^{A} = - \frac{1}{4} \partial_{x}^2 \pm q^2 \pm q_{x} I_{-} q,
\end{align}
to give
\begin{align}
    \int_{-\infty}^{\infty} \left\{ q_{t} + \Theta(L_{\pm}^{A}) q_{x} \right\} \mu_{\pm} dx = 0,
\end{align}
which implies
\begin{align}
    q_{t} + \Theta(L_{\pm}^{A}) q_{x} = 0. \label{eqn:scalar_general_evolution_equation}
\end{align}
This defines the family of nonlinear evolution equations associated to the scalar scattering system in equations (\ref{eqn:scalar_scattering_1}) and (\ref{eqn:scalar_scattering_2}). Notice that if we take $\Theta(L_{\pm}^{A}) = - 4 L_{\pm}^{A}$, equation (\ref{eqn:scalar_general_evolution_equation}) gives mKdV
\begin{align}
    q_{t} + q_{xxx} \mp 6 q^2 q_{x} = 0.
\end{align}
We can relate the operator $\Theta$ directly to the dispersion relation of the linearization of equation \ref{eqn:scalar_general_evolution_equation}. As $L_{\pm}^{A} \to -\frac{1}{4} \partial_{x}^2$ implies $\Theta(L_{\pm}^{A}) \to \Theta(-\frac{1}{4} \partial_{x}^2)$, this linearization is
\begin{align}
    q_{t} + \Theta\left(-\partial_{x}^2/4\right) q_{x} = 0.
\end{align}
Putting $q = e^{i ( k x - \omega(k) t)}$ gives
\begin{align}
    \Theta(k^2) = \frac{\omega(2 k)}{2 k}. \label{eqn:scalar_dispersion_relation}
\end{align}
Therefore, using our definitions of linear fmKdV and linear fsineG and fsinhG, with dispersion relations $\omega(k) = - k^3 |k|^{2\epsilon}$ and $\omega(k) = |k|^{2\epsilon}/k$ where $0 < \epsilon < 1$, respectively, we have $\Theta(L_{\pm}^{A}) = - 4 L_{\pm}^{A} |4 L_{\pm}^{A}|^{\epsilon}$, $\Theta(L_{-}^{A}) = \frac{|4 L_{-}^{A}|^{\epsilon}}{4 L_{-}}$, and $\Theta(L_{+}^{A}) = \frac{|4 L_{+}^{A}|^{\epsilon}}{4 L_{+}}$, respectively (recall that fmKdV has $r = \pm q$ while fsineG and fsinhG have $r = -q$ and $r = +q$, respectively). Therefore, we can write fmKdV, fsineG, and fsinhG as
\begin{align}
    q_{t} - 4 L_{\pm}^{A} |2 L_{\pm}^{A}|^{\epsilon} q_{x} &= 0, \label{eqn:fmKdV} \\
    q_{t} + \frac{|4L_{-}^{A}|^\epsilon}{4 L_{-}^{A}} q_{x} = 0, ~~ u_{tx}& + \frac{|4L_{-}^{A}|^\epsilon}{4 L_{-}^{A}} u_{xx}, \label{eqn:fsineG} \\
    q_{t} + \frac{|4L_{+}^{A}|^\epsilon}{4 L_{+}^{A}} q_{x} = 0, ~~ u_{tx}& + \frac{|4L_{+}^{A}|^\epsilon}{4 L_{+}^{A}} u_{xx}. \label{eqn:fsinhG}
\end{align}
Notice that as both $L_{\pm} \to - \partial_{x}^{2}/4$, in the linear limit, fsineG and fsinhG both converge to (\ref{eqn:linear_SG}). Currently, the meaning of $|L_{\pm}^{A}|^{\epsilon}$ is not clear; it will be defined in the next section using a spectral expansion in terms of the squared eigenfunctions $\mu_{\pm}$ and $\nu_{\pm}$.

\subsection{Completeness of Squared Scalar Eigenfunctions}

In \cite{Kaup76} it was shown that the eigenfunctions $\mathbf{\Psi}$ and $\mathbf{\Psi}^{A}$, equations (\ref{eqn:bold_psi_definition}) and (\ref{eqn:bold_psi_A_definition}), of the AKNS system are complete. Specifically, for a sufficiently smooth and decaying vector-valued function $\mathbf{v}(x) = \left( v^{(1)}(x),v^{(2)}(x) \right)^{T}$, we have
\begin{align}
    \mathbf{v}(x) &= \sum_{n = 1}^{2}\int_{\Gamma_{\infty}^{(n)}} dk f_{n}(k) \int_{-\infty}^{\infty} dy\,\mathbf{G}_{n}(x,y,k)\mathbf{v}(y),  \label{eqn:AKNS_completness_relation} \\
    \mathbf{G}_{1}(x,y,k) &= \mathbf{\Psi}(x,k) \mathbf{\Psi}^{A}(y,k)^{T}, \quad f_{1}(k) = -\tau^2(k)/\pi, \notag \\
    \mathbf{G}_{2}(x,y,k) &= \overline{\mathbf{\Psi}}(x,k) \overline{\mathbf{\Psi}}^{A}(y,k)^{T}, \quad f_{2}(k) = \overline{\tau}^2(k)/\pi, \notag
 \end{align}
where $\Gamma_{R}^{(1)}$ ($\Gamma_{R}^{(2)}$) is the semicircular contour in the upper (lower) half plane evaluated from $-R$ to $+R$ and $\tau(k)$ and $\overline{\tau}(k)$ are transmission coefficients defined in equations (\ref{eqn:AKNS_scattering_data_phi}) and (\ref{eqn:AKNS_scattering_data_psi}). Notice that $\mathbf{G}_{n}$, $n = 1,2$ are $2\times2$ matrices. However, here we need to use the adjoint completeness relation, which may be found directly from (\ref{eqn:AKNS_completness_relation}) using the inner product $(\mathbf{u},\mathbf{v}) := \int_{-\infty}^{\infty} \mathbf{u}(x)^{T} \mathbf{v}(x) dx$ where $\mathbf{u}$, $\mathbf{v}$ are $2\times1$ column vectors. To do this, we expand $\mathbf{v}$ using the above completeness relation, and then exchange the order of integration to find an expansion for $\mathbf{u}$. This procedure gives us
\begin{align}
    \mathbf{v}(x) &= \sum_{n = 1}^{2}\int_{\Gamma_{\infty}^{(n)}} dk f_{n}(k) \int_{-\infty}^{\infty} dy\,\mathbf{G}_{n}^{A}(x,y,k)\mathbf{v}(y),  \label{eqn:AKNS_adjoint_completness_relation} \\
    \mathbf{G}_{1}^{A}(x,y,k) &= \mathbf{\Psi}^{A}(x,k) \mathbf{\Psi}(y,k)^{T}, \quad f_{1}(k) = -\tau^2(k)/\pi, \notag \\
    \mathbf{G}_{2}^{A}(x,y,k) &= \overline{\mathbf{\Psi}}^{A}(x,k) \overline{\mathbf{\Psi}}(y,k)^{T}, \quad f_{2}(k) = \overline{\tau}^2(k)/\pi. \notag
 \end{align}
However, when $r = \pm q$, the symmetry reductions in equations (\ref{eqn:scalar_symmetry_reduction_eigenfunctions}) and (\ref{eqn:scalar_symmetry_reduction_data}) give
\begin{align}
    \overline{\mathbf{\Psi}}(k) = \sigma_{+} \mathbf{\Psi}(-k), ~~ \overline{\mathbf{\Psi}}^{A}(k) = - \sigma_{+} \mathbf{\Psi}^{A}(-k).
\end{align}
Therefore, the adjoint completness relation in (\ref{eqn:AKNS_adjoint_completness_relation}) reduces to
\begin{align}
    \mathbf{v}(x) = &- \int_{\Gamma_{\infty}^{(1)}} dk \frac{\tau^2(k)}{\pi} \int_{-\infty}^{\infty} dy \mathbf{\Psi}^{A}(x,k) \mathbf{\Psi}(y,k)^{T} \mathbf{v}(y) \\
    &- \int_{\Gamma_{\infty}^{(2)}} dk \frac{\tau^2(-k)}{\pi} \int_{-\infty}^{\infty} dy \sigma_{+} \mathbf{\Psi}^{A}(x,-k) \mathbf{\Psi}(y,-k)^{T} \sigma_{+} \mathbf{v}(y).
\end{align}
The second integral may be rewritten with the substitution $\xi = - k$ as
\begin{align}
    - \int_{\Gamma_{\infty}^{(1)}} d\xi \frac{\tau^2(\xi)}{\pi} \int_{-\infty}^{\infty} dy \sigma_{+} \mathbf{\Psi}^{A}(x,\xi) \mathbf{\Psi}(y,\xi)^{T} \sigma_{+} \mathbf{v}(y).
\end{align}
Therefore, we have
\begin{align}
    \mathbf{v}(x) = \int_{\Gamma_{\infty}^{(1)}} d\xi \frac{\tau^2(\xi)}{\pi} \int_{-\infty}^{\infty} dy \left[ \mathbf{\Psi}^{A}(x,k) \mathbf{\Psi}(y,k)^T - \sigma_{+} \mathbf{\Psi}^{A}(x,k) \mathbf{\Psi}(y,k)^T \sigma_{+} \right] \mathbf{v}(y).
\end{align}
If we put $\mathbf{v}(x) = \mathbf{h}(x) = \left(h(x),\mp h(x)\right)^{T}$, we can reduce the completeness relation to
\begin{align}
    h(x) = \mp \int_{\Gamma_{\infty}^{(1)}} dk \int_{-\infty}^{\infty} dy \, g_{\pm}(x,y,k) h(y), \label{eqn:scalar_completness}
\end{align}
for the scalar function $h(x)$ where
\begin{align}
    g_{\pm}(x,y,k) = \frac{\tau^2(k)}{\pi} \nu_{\pm}(x,k) \mu_{\pm}(y,k), \label{eqn:scalar_kernel_g}
\end{align}
with $\nu_{\pm}(x,k) = (\phi^{(1)}(x,k))^{2} \pm (\phi^{(2)}(x,k))^{2}$ and $\mu_{\pm}(x,k) = ( \psi^{(1)}(x,k))^{2} \mp ( \psi^{(2)}(x,k))^{2}$. Then, the action of the operator $\Theta(L_{\pm}^{A})$ on the function $h(x)$ may be written as
\begin{align}
    \Theta(L_{\pm}^{A}) h(x) = \mp \int_{\Gamma_{\infty}^{(1)}} dk \Theta(k^2) \int_{-\infty}^{\infty} dy \,  g_{\pm}(x,y,k) h(y), \label{eqn:L_pm_spectral_expansion}
\end{align}
and the famility of nonlinear evolution equations in equation (\ref{eqn:scalar_general_evolution_equation}) becomes
\begin{align}
    q_{t} \mp \int_{\Gamma_{\infty}^{(1)}} dk \Theta(k^2) \int_{-\infty}^{\infty} dy \,  g_{\pm}(x,y,k) \partial_{y} q(y) = 0. \label{eqn:scalar_general_evolution_equation_expansion} 
\end{align}
In particular, fmKdV can be represented as
\begin{align}
    q_{t} \mp \int_{\Gamma_{\infty}^{(1)}} dk |2 k|^{2\epsilon} \int_{-\infty}^{\infty} dy \,  g_{\pm}(x,y,k) \left[ q_{yyy} \mp 6 q^2 \right] = 0, \label{eqn:fmKdV_expansion} 
\end{align}
and fsineG and fsinhG are given by
\begin{align}
    q_{t} &\mp \int_{\Gamma_{\infty}^{(1)}} dk |2 k|^{2(1 - \epsilon)} \int_{-\infty}^{\infty} dy \,  g_{-}(x,y,k) \partial_{y} q(y) = 0, \label{eqn:fsineG_expansion} \\
    q_{t} &\mp \int_{\Gamma_{\infty}^{(1)}} dk |2 k|^{2(1 - \epsilon)} \int_{-\infty}^{\infty} dy \,  g_{+}(x,y,k) \partial_{y} q(y) = 0. \label{eqn:fsinhG_expansion} 
\end{align}
Because the $k$ integral in equation (\ref{eqn:scalar_completness}) is over the semicircle in the upper half plane, it can be expressed instead in terms of an integral along the real line and a sum over the residues using contour integration. This is a useful representation because it explicitly separates the continuous and discrete spectra, where the latter corresponds to bound states at $k = k_{j}$, $j = 1, 2, ..., J$. Using the closed contour composed of $\Gamma^{(1)}$ and an integral along the real line from $\infty$ to $-\infty$, we can write
\begin{align}
    \Theta(L_{\pm}^{A})h(x) = &\mp \int_{-\infty}^{\infty} dk \Theta(k^2) \int_{-\infty}^{\infty} dy \, g_{\pm}(x,y,k) h(y),  \\
    &\pm 2 \pi i \sum_{j = 1}^{J} \text{Res}\left( \Theta(k^2) \int_{-\infty}^{\infty} dy \, g_{\pm}(x,y,k) h(y), k_{j} \right). \notag
\end{align}
Because $\nu_{\pm}$ and $\mu_{\pm}$ are all analytic in the upper half plane, the only residues come from the poles of $\tau^2$, or zeros of $a^2$. These occur at $k_{j} = \xi_{j} + i \eta_{j}$, $j = 1, 2, ..., J$ and are assumed to be simple, meaning $\tau^2$ has a double pole at $k_{j}$. Therefore, we can compute the residue at $k_{j}$ as
\begin{align}
    \text{Res}&\left( \int_{-\infty}^{\infty} dy \, g_{\pm}(x,y,k) h(y), k_{j} \right) = \lim_{k\to k_{j}} \frac{\partial}{\partial k} \left[ (k - k_{j})^2 \Theta(k^2) \int_{-\infty}^{\infty} dy \, g_{\pm}(x,y,k) h(y) \right], \notag \\
    &= \frac{\Theta(k_{j}^{2})}{\pi (a_{j}')^2} \int_{-\infty}^{\infty} dy \left\{ \partial_{k} \nu_{\pm}(x,k) \mu_{\pm}(y,k) + \nu_{\pm}(x,k) \partial_{k} \mu_{\pm}(x,k) \right\}_{k = k_{j}} h(y), \\
    &+ \left( \frac{2 k_{j} \Theta'(k_{j}^2)}{\pi (a_{j}')^2}  - \frac{a_{j}'' \Theta(k_{j}^2)}{\pi (a_{j}')^3} \right) \int_{-\infty}^{\infty} dy \, \nu_{\pm}(x,k_{j}) \mu_{\pm}(y,k_{j}) h(y). \notag
\end{align}
Defining
\begin{align}
    g_{\pm,j}^{(1)}(x,y) &= \frac{2 i}{(a_{j}')^2} \left\{ \partial_{k} \nu_{\pm}(x,k) \mu_{\pm}(y,k) + \nu_{\pm}(x,k) \partial_{k} \mu_{\pm}(y,k) \right\}_{k = k_{j}}, \\
    g_{\pm,j}^{(2)}(x,y) &= \frac{2 i} {(a_{j}')^2} \nu_{\pm}(x,k_{j}) \mu_{\pm}(y,k_{j}), \\
    g_{\pm,j}^{(3)}(x,y) &= - \frac{2 i a_{j}''}{(a_{j}')^3} \nu_{\pm}(x,k_{j}) \mu_{\pm}(y,k_{j}),
\end{align}
we have
\begin{align}
    h(x) = &\mp \int_{-\infty}^{\infty} dk \Theta(k^2) \int_{-\infty}^{\infty} dy \, g_{\pm}(x,y,k) h(y), \label{eqn:L_pm_spectral_expansion_continuous_discrete_split} \\
    &\pm \sum_{j = 1}^{J} \int_{-\infty}^{\infty} dy \left\{ \Theta(k_{j}^2) g_{\pm,j}^{(1)}(x,y) + 2 k_{j} \Theta'(k_{j}^{2}) g^{(2)}_{\pm,j}(x,y) + \Theta(k_{j}^2) g_{\pm,j}^{(3)}(x,y) \right\} h(y). \notag
\end{align}
Notice that if we take $\Theta(k^2) = 1$, then we have the identity in (\ref{eqn:scalar_completness}) written with continuous and discrete spectra separated.

\section{The IST for Fractional Modified KdV, SineG, and SinhG}

Solving nonlinear evolution equations with the IST is analogous to solving linear evolution equations with Fourier transforms. To solve linear problems, the Fourier transform is taken to map the problem into Fourier space where the time evolution is described by a simple set of differential equations. These equations are then solved to give the solution at any time $t$ in Fourier space. Finally, the solution is mapped back to physical space using the inverse Fourier transform, which amounts to evaluating an integral. Mapping the initial condition into scattering space via direct scattering is analogous to taking the Fourier transform, time evolution in scattering space is nearly identical to that in Fourier space, and inverse scattering maps the solution to the nonlinear problem back into physical space just as the inverse Fourier transform does. The major difference between Fourier transforms and the IST is that performing integrals for the Fourier transform and inverse Fourier transform is replaced by solving linear integral equations for direct scattering and inverse scattering. In the following we, outline direct scattering, time evolution, and inverse scattering for the scalar scattering system.

\subsection{Direct Scattering}

To solve the nonlinear evolution equation, equation (\ref{eqn:scalar_general_evolution_equation}), by the inverse scattering transform, we first map the initial condition into scattering space; this is analogous to taking the Fourier transform of a linear partial differential equation. This process involves analyzing linear integral equations for the eigenfunctions, determining their analytic properties, and then obtaining the scattering data using Wronskian relations.

Eigenfunctions of the scalar scattering problem are precisely $\boldsymbol\phi$ and $\boldsymbol\psi$ of the AKNS system with the symmetry reduction in equations (\ref{eqn:scalar_symmetry_reduction_eigenfunctions}) and (\ref{eqn:scalar_symmetry_reduction_data}). They are solutions to equations (\ref{eqn:scalar_scattering_1}) and (\ref{eqn:scalar_scattering_2}) subject to the boundary conditions in equations (\ref{eqn:AKNS_asymptotic_phi}) and (\ref{eqn:AKNS_asymptotic_psi}) with the scattering data defined by equation (\ref{eqn:AKNS_scattering_data_phi}). It is convenient to express the scattering functions in terms of \emph{Jost solutions} by taking
\begin{alignat}{2}
    \mathbf{M}(x,k,t) &= e^{i k x} \boldsymbol\phi(x,k,t), \quad &&\mathbf{N}(x,k,t) = e^{- i k x} \boldsymbol\psi(x,k,t). \label{eqn:jost_definition} \\
    \overline{\mathbf{M}}(x,k,t) &= \sigma^{-1}_{\pm} \mathbf{M}(x,-k,t), \quad &&\overline{\mathbf{N}}(x,k,t) = \sigma_{\pm} \mathbf{N}(x,-k,t), \label{eqn:jost_definition_bar}
\end{alignat}
where the symmetry reductions for $r = \pm q$ are in terms of 
\begin{align}
    \sigma_{\pm} = \begin{pmatrix}0&1\\\pm1&0\end{pmatrix}, \quad \sigma^{-1}_{\pm} = \begin{pmatrix}0&\pm1\\\ 1&0\end{pmatrix}.
\end{align}
Then, the boundary conditions become constant
\begin{align}
    \mathbf{M}(x,k,t) &\sim \begin{pmatrix}
    1 \\
    0
    \end{pmatrix},~~ x\to-\infty,~~~\mathbf{N}(x,k,t) \sim \begin{pmatrix}
    0 \\
    1
    \end{pmatrix}, ~~x\to\infty, \label{eqn:scalar_asymptotic_jost}
\end{align}
and the scattering equation, where either $\mathbf{M}$ or $\mathbf{N}$ is represented generically as $\boldsymbol{\chi} = \boldsymbol{\chi}(x,k,t)$, becomes
\begin{align}
    \partial_{x} \boldsymbol{\chi} = i k \mathbf{B} \boldsymbol{\chi} + \mathbf{Q} \boldsymbol{\chi}, \label{eqn:scalar_scattering_jost}
\end{align}
where
\begin{align}
    \mathbf{B} = \begin{pmatrix}0&0\\0&2\end{pmatrix}, \quad \mathbf{Q} = \begin{pmatrix}0&q\\\pm q&0\end{pmatrix}. \quad
\end{align}
This differential equation can be converted to an integral equation for $\mathbf{M}$ and $\mathbf{N}$, cf. \cite{Ablowitz2011},
\begin{align}
    \mathbf{M}(x,k,t) &= \begin{pmatrix}1\\0\end{pmatrix} + \int_{-\infty}^{\infty} \mathbf{G}(x - \xi,k,t) \mathbf{Q}(\xi,t) \mathbf{M}(\xi,k,t) \, d\xi, \label{eqn:direct_scattering_M} \\
    \mathbf{N}(x,k,t) &= \begin{pmatrix}0\\1\end{pmatrix} + \int_{-\infty}^{\infty} \overline{\mathbf{G}}(x - \xi,k,t) \mathbf{Q}(\xi,t) \mathbf{N}(\xi,k,t) \, d\xi, \label{eqn:direct_scattering_N}
\end{align}
where
\begin{align}
    \mathbf{G}(x,k,t) &=  \theta(x) \begin{pmatrix} 1&0\\0&e^{2 i k x}\end{pmatrix}, \\
    \overline{\mathbf{G}}(x,k,t) &=  -\theta(-x) \begin{pmatrix} e^{- 2 i k x}&0\\0&1\end{pmatrix}.
\end{align}
So long as $q,r\in L^{1}(\mathbb{R})$, these Volterra integral equations have absolutely and uniformly convergent Neumann series in the upper half $k$-plane \cite{Ablowitz3}. Therefore, the functions $\mathbf{M}$ and $\mathbf{N}$ are analytic functions of $k$ for $\text{Im}\,k>0$ and continuous for $\text{Im}\,k\geq0$. This also implies that $\overline{\mathbf{M}}$ and $\overline{\mathbf{N}}$ are analytic for $\text{Im}\,k<0$ and continuous for $\text{Im}\,k\leq0$ from their relations in equation (\ref{eqn:jost_definition_bar}). Using these integral equations and the intial condition at $t = 0$, the Jost solutions $\mathbf{M}$ and $\mathbf{N}$ can be constructed at $t = 0$ and, subsequently, the scattering functions from the relations in equation (\ref{eqn:jost_definition}). Then, the initial scattering data may be derived from the Wronskian relations in equations (\ref{eqn:AKNS_wronskian_relations_a}) and (\ref{eqn:AKNS_wronskian_relations_b}).

We will also need the asymptotic properties of $\mathbf{N}$ and $\mathbf{M}$ to reconstruct the solution in inverse scattering; so, expanding equations (\ref{eqn:direct_scattering_M}) and (\ref{eqn:direct_scattering_N}) in large $k$, after integrating by parts, we have
\begin{align}
    \mathbf{M}(x,k,t) &= \begin{pmatrix} 1 - \frac{1}{2 i k} \int_{-\infty}^{x} q(\xi,t) r(\xi,t) \, d\xi \\- \frac{1}{2 i k } r(x,t) \end{pmatrix} + \mathcal{O}(k^{-2}) \label{eqn:direct_scattering_asymptotics_M} \\
    \mathbf{N}(x,k,t) &= \begin{pmatrix} \frac{1}{2 i k} q(x,t) \\ 1 - \frac{1}{2 i k} \int_{x}^{\infty} q(\xi,t) r(\xi,t) \, d\xi \end{pmatrix} + \mathcal{O}(k^{-2}) \label{eqn:direct_scattering_asymptotics_N}
\end{align}

\subsection{Time Evolution}

After the initial condition is projected into scattering space by reconstructing the scattering functions and scattering data, the data is evolved in time by solving a simple set of ordinary differential equations. The scattering functions evolve in time according to
\begin{align}
    \mathbf{v}_{t} = \begin{pmatrix} A&B\\C&-A \end{pmatrix} \mathbf{v} \label{eqn:AKNS_scattering_function_time_evolution}
\end{align}
where $A$, $B$, $C$ are functions of $x$, $k$, $t$ which cannot be represented generally. However, their asymptotic properties can be used to characterize the time evolution of the scattering data \cite{Ablowitz2011} as
\begin{alignat}{2}
    a(k,t) &= a(k,0), ~~ &&b(k,t) = b(k,0) e^{- 2 i k \Theta(k^2) t}, \label{eqn:scalar_scattering_data_time_evolution_a_b} \\
    \rho(k,t) &= \rho(k,0) e^{- 2 i k \Theta(k^2) t}, ~~ &&C_j(t) = C_j(0)  e^{- 2 i k_{j} \Theta(k_{j}^2) t}, \label{eqn:scalar_scattering_data_time_evolution_rho_c}
\end{alignat}
for $j = 1, 2, ..., J$. We recall that $\Theta(k^2)$ is related to the linear dispersion relation as in equation (\ref{eqn:scalar_dispersion_relation}). To characterize the spectral expansion in equation (\ref{eqn:scalar_general_evolution_equation_expansion}), we need to be able to compute the time evolution of the scattering functions $\boldsymbol\psi$ and $\boldsymbol\phi$. Although equation (\ref{eqn:AKNS_scattering_function_time_evolution}) does not give this in a simple way, the scattering functions can be evolved in time using inverse scattering, which is discussed next.

\subsection{Inverse Scattering}

Inverse scattering allows the construction of the solution to the nonlinear evolution equation $q(x,t)$ and the scattering functions $\boldsymbol\psi(x,k,t)$ and $\boldsymbol\phi(x,k,t)$ from the scattering data, $a(k,t)$ and $b(k,t)$ obtained from equation (\ref{eqn:scalar_scattering_data_time_evolution_a_b}). For the Jost solutions (\ref{eqn:jost_definition}), we can write the scattering data in equation (\ref{eqn:AKNS_scattering_data_phi}), using (\ref{eqn:AKNS_tau_rho}), as
\begin{align}
    \boldsymbol\mu(x,k,t) = \overline{\mathbf{N}}(x,k,t) + \rho(k,t)e^{2 i k x} \mathbf{N}(x,k,t), \quad \boldsymbol\mu(x,k,t) = \mathbf{M}(x,k,t)/a(k,t), \label{eqn:riemann_hilbert_problem}
\end{align}
where $\mathbf{M}$ is analytic in the upper half plane, $\boldsymbol\mu$ is meromorphic with simple poles at the zeros of $a$, and $\overline{\mathbf{N}}$ is analytic in the lower half plane ($\rho$ is not analytic in general). Therefore, equation (\ref{eqn:riemann_hilbert_problem}) defines the ``jump'' condition of a Riemann-Hilbert problem which we will transform onto an integral equation for $\mathbf{N}$. This equation will allow the construction of the scattering function $\boldsymbol\psi$ and the solution $q(x,t)$. We will also outline how the same method can be used to derive an equation for $\mathbf{M}$.

We assume that $a$ has simple zeros, and hence $\boldsymbol{\mu}$ has simple poles in the upper half plane at $k_{j}$ for $j = 1, 2, ..., J$ with no zeros along the real line. Then, as $\boldsymbol\mu$ has only simple poles, we can represent it as
\begin{align}
    \boldsymbol\mu(x,k,t) = \mathbf{h}(x,k,t) + \sum_{j = 1}^{J} \frac{\mathbf{A}_{j}(x,t)}{k - k_{j}},
\end{align}
where $\mathbf{h}$ is analytic in $k$ for $\text{Im}\, k>0$. By integrating in a small neighborhood around each $k_{j}$, and using equation (\ref{eqn:riemann_hilbert_problem}), we find that $\mathbf{A}_{j}$ is given by
\begin{align}
    \mathbf{A}_{j}(x,t) = C_{j}(t) e^{2 i k_{j} x} \mathbf{N}_{j}(x,t), \text{ for } j = 1, 2, ..., J, \label{eqn:A_mu_residue}
\end{align}
where $C_{j}(t) = b_{j}(t)/a_{j}'(t)$ and $\mathbf{N}_{j}(x,t) = \mathbf{N}(x,k_{j},t)$. We then define the projection operators
\begin{align}
    P^{\pm}[f](k) = \frac{1}{2 \pi i} \int_{-\infty}^{\infty} \frac{f(\xi)}{\xi - (k \pm i 0)} d\xi. \label{eqn:projection_operator_definition}
\end{align}
If $f_{+}$ ($f_{-}$) is analytic in the upper (lower) half plane and $f_{\pm}(k) \to 0$ as $|k|\to\infty$ for $\text{Im}\,k>0$ ($\text{Im}\,k<0$), then
\begin{align}
    P^{\pm}[f_{\pm}] = \pm f_{\pm}, \quad P^{\pm}[f_{\mp}] = 0. \label{eqn:projection_operator_action}
\end{align}
Taking equation (\ref{eqn:riemann_hilbert_problem}) and subtracting $(1,0)^{T}$ and the simple poles of $\boldsymbol\mu$, we have
\begin{align}
    \mathbf{h}(x,k,t) - \begin{pmatrix}1\\0\end{pmatrix} = \overline{\mathbf{N}}(x,k,t) - \begin{pmatrix}1\\0\end{pmatrix} -  \sum_{j = 1}^{J} \frac{\mathbf{A}_{j}(x,t)}{k - k_{j}} + \rho(k,t) e^{2 i k x} \mathbf{N}(x,k,t). \label{eqn:pre_projection_riemann_hilbert}
\end{align}
The left side is analytic in the upper half plane and approaches zero as $|k|\to\infty$; $\overline{\mathbf{N}} - (1,0)^{T}$ is also analytic in the lower half plane and vanishes asymptotically. Therefore, applying $P^{-}$ to (\ref{eqn:pre_projection_riemann_hilbert}) gives
\begin{align}
    \overline{\mathbf{N}}(x,k,t) = \begin{pmatrix}1\\0\end{pmatrix} +  \sum_{j = 1}^{J} \frac{\mathbf{A}_{j}(x,t)}{k - k_{j}} + \frac{1}{2 \pi i} \int_{-\infty}^{\infty} \frac{\rho(\xi,t) e^{2 i \xi x}}{\xi - (k - i 0)} \mathbf{N}(x,\xi,t) \, d\xi.
\end{align}
Noting that $\overline{\mathbf{N}}(x,k,t) = \sigma\mathbf{N}(x,-k,t)$ and using the expression for $\mathbf{A}_{j}$ in equation (\ref{eqn:A_mu_residue}), we find the integral equation for $\mathbf{N}$
\begin{align}
    \mathbf{N}(x,k,t) = \begin{pmatrix}0\\1\end{pmatrix} -  \sum_{j = 1}^{J} \frac{C_{j}(t) e^{2 i k_{j} x}}{k + k_{j}} \sigma^{-1} \mathbf{N}_{j}(x,t) + \frac{1}{2 \pi i} \int_{-\infty}^{\infty} \frac{\rho(\xi,t) e^{2 i \xi x}}{\xi + k + i 0} \sigma^{-1} \mathbf{N}(x,\xi,t) \, d\xi. \label{eqn:N_inverse_scattering_integral_equation}
\end{align}
Evaluating this at $k_{\ell}$ for $\ell = 1,2,...,J$, we obtain an equation for $\mathbf{N}_{\ell}(x,t)$.
\begin{align}
    \mathbf{N}_{\ell}(x,t) = \begin{pmatrix}0\\1\end{pmatrix} -  \sum_{j = 1}^{J} \frac{C_{j}(t) e^{2 i k_{j} x}}{k_{\ell} + k_{j}} \sigma^{-1} \mathbf{N}_{j}(x,t) + \frac{1}{2 \pi i} \int_{-\infty}^{\infty} \frac{\rho(\xi,t) e^{2 i \xi x}}{\xi + k_{\ell}} \sigma^{-1} \mathbf{N}(x,\xi,t) \, d\xi. \label{eqn:N_ell_inverse_scattering_integral_equation}
\end{align}
We can similarly show that $\mathbf{M}$ and $\mathbf{M}_{j}$ solve 
\begin{align}
    \mathbf{M}(x,k,t) = \begin{pmatrix}1\\0\end{pmatrix} + \sum_{j = 1}^{J} \frac{\tilde{C}_{j}(t) e^{-2 i k_{j} x}}{k + k_{j}} \sigma\mathbf{M}_{j}(x,t) - \frac{1}{2 \pi i} \int_{-\infty}^{\infty} \frac{\tilde{\rho}(\xi,t) e^{-2 i \xi x}}{\xi + k + i 0} \sigma \mathbf{M}(x,\xi,t) \, d\xi, \label{eqn:M_inverse_scattering_integral_equation} \\
    \mathbf{M}_{\ell}(x,t) = \begin{pmatrix}1\\0\end{pmatrix} +  \sum_{j = 1}^{J} \frac{\tilde{C}_{j}(t) e^{-2 i k_{j} x}}{k_{\ell} + k_{j}} \sigma \mathbf{M}_{j}(x,t) - \frac{1}{2 \pi i} \int_{-\infty}^{\infty} \frac{\tilde{\rho}(\xi,t) e^{-2 i \xi x}}{\xi + k_{\ell}} \sigma \mathbf{M}(x,\xi,t) \, d\xi. \label{eqn:M_ell_inverse_scattering_integral_equation}
\end{align}
These equations can then be expanded for large $k$ and, by comparing these expansions to those for the direct scattering problem in equations (\ref{eqn:direct_scattering_asymptotics_M}) and (\ref{eqn:direct_scattering_asymptotics_N}), we can recover the solution at any time $q(x,t)$ from $\mathbf{N}(x,k,t)$ as
\begin{align}
    q(x,t) = \mp 2 i \sum_{j = 1}^{J} e^{2 i k_{j} x} C_{j}(t) N_{j}^{(2)}(x,t) \pm \frac{1}{\pi} \int_{-\infty}^{\infty} \rho(\xi,t) e^{2 i \xi x} N^{(2)}(x,\xi,t) \, d\xi. \label{eqn:inverse_scattering_solution_reconstruction}
\end{align}
Notice that just as the spectral expansion of the $\Theta(L_{\pm}^{A})$ split into discrete and continuous spectra in equation (\ref{eqn:L_pm_spectral_expansion_continuous_discrete_split}), the solution $q$ is composed of a sum over discrete contributions and an integral over continuous contributions. These equations can be converted into the following GLM type integral equations \cite{AKNS}
\begin{align}
    \mathbf{K}(x,y;t) \pm \begin{pmatrix} 1 \\ 0 \end{pmatrix} F(x+y;t) + \int_{x}^{\infty}  \sigma^{-1} \mathbf{K}(x,s;t) F(s+y;t) \, ds = 0, \label{eqn:inverse_scattering_GLM_integral_equation_K} \\
    \mathbf{L}(x,y;t) \pm \begin{pmatrix} 0 \\ 1 \end{pmatrix} G(x+y;t) + \int_{-\infty}^{x}  \sigma^{-1} \mathbf{L}(x,s;t) G(s+y;t) \, ds = 0, \label{eqn:inverse_scattering_GLM_integral_equation_L}
\end{align}
where
\begin{align}
    F(x;t) &=
    \frac{1}{2\pi} \int_{-\infty}^{\infty} \rho(\xi,t) e^{+i\xi x} d\xi - i\sum_{j=1}^{J}C_{j}(t) e^{ik_{j}x},\\
    G(x;t) &=
    \frac{1}{2\pi} \int_{-\infty}^{\infty} \tilde{\rho}(\xi,t) e^{-i\xi x} d\xi - i\sum_{j=1}^{J}\tilde{C}_{j}(t) e^{ik_{j}x},
\end{align}
and the Jost eigenfunctions are related to the triangular kernel by
\begin{align}
    \mathbf{N}(x;k,t) &= \begin{pmatrix} 0 \\ 1 \end{pmatrix} + \int_{x}^{\infty} \mathbf{K}(x,s;t) e^{- i k (x - s)} ds, ~ \text{Im}\,k>0, \\
    \mathbf{M}(x;k,t) &= \begin{pmatrix} 1 \\ 0 \end{pmatrix} - \int_{-\infty}^{x} \mathbf{L}(x,s;t) e^{+i k (x - s)} \, ds, ~ \text{Im}\,k>0.
\end{align}
The solution of the nonlinear scalar equation can then be obtained from
\begin{align}
    q(x,t) = -2 K^{(1)}(x,x;t),  \label{eqn:inverse_scattering_GLM_solution_reconstruction}
\end{align}
where $K^{(1)}$ denotes the $1$st component of the vector $\mathbf{K}$.

\section{The One Soliton Solution}

Pure soliton solutions of the scalar general evolution equation (\ref{eqn:scalar_general_evolution_equation}) are reflectionless, i.e., $\rho(k,t) = 0$ on the real line. They are also bound states corresponding to the discrete eigenvalues at the zeros of $a$. We note that soliton solutions for $r = q$ do not exist when we assume  $q$ and $r$ vanish sufficiently rapidly as $|x|\to\infty$.  Because of this, we will only consider $r = -q$ for the remainder of our study. For a given initial condition, the number of discrete eigenvalues $k_{j} = \xi_{j} + i \eta_{j}$ $j = 1, 2, ..., J$ gives the number of solitons. The general $J$-soliton solution can be reduced to solving a linear algebraic system. For simplicity
we consider the one-soliton solution, $J = 1$, although the argument we lay out can be used also for larger $J$. We find the one-soliton solution by constructing $\mathbf{N}(x,k,t)$ from equation (\ref{eqn:N_inverse_scattering_integral_equation}) and (\ref{eqn:N_ell_inverse_scattering_integral_equation}) and then recovering $q(x,t)$ using (\ref{eqn:inverse_scattering_solution_reconstruction}). We will then explicitly verify that this solution is in fact a solution of the general evolution equation characterized in physical space by the completeness relation, equation (\ref{eqn:scalar_general_evolution_equation_expansion}), using complex variable methods. This will require that we construct $\mathbf{N}(x,k,t)$ and $\mathbf{M}(x,k,t)$ (thereby $\mu_{\pm}(x,k,t)$, $\nu_{\pm}(x,k,t)$ and $\tau(k,t)$).

\subsection{Deriving the One Soliton from Inverse Scattering}

Putting $\rho = 0$ and $J = 1$ into equation (\ref{eqn:N_ell_inverse_scattering_integral_equation}) and taking $k_{1} = i \eta$ and $C_{1}(0) = -2 i \eta e^{2 \eta x_{0}}$ such that $C_{1}(t) = -2 i \eta e^{2 \eta x_{0} + 2 \eta \Theta(-\eta^2) t}$ gives an algebraic equation for $\mathbf{N}_{1}(x,t)$
\begin{align}
    \mathbf{N}_{1}(x,t) = \begin{pmatrix}0\\1\end{pmatrix} + e^{- z_{t}(x)} \sigma^{-1} \mathbf{N}_{1}(x,t), \label{eqn:one_soliton_N_ell_equation}
\end{align}
where $z_{t}(x) = 2 \eta (x - x_{0}) - 2 \eta \Theta(- \eta^2) t$. The eigenvalue $k_{1}$ is imaginary because for $r = - q$, all discrete eigenvalues come in pairs $\{k_{j}, - k_{j}^{*}\}_{j = 1}^{J}$. Solving this gives
\begin{align}
    N_{1}^{(1)}(x,t) = -\frac{1}{2} \text{sech}\{ z_{t}(x)\}, ~~ 
    N_{1}^{(2)}(x,t) = \frac{1}{2}\left(1 + \text{tanh}\{ z_{t}(x)\}\right),
\end{align}
and then putting these components into equation (\ref{eqn:inverse_scattering_solution_reconstruction}) yields
\begin{align}
    q(x,t) =  2 \eta\, \text{sech}\{z_{t}(x)\}. \label{eqn:one_soliton_solution}
\end{align}

\subsection{Verifying the One Soliton}

To verify that the one soliton given in equation is truly a solution to the general evolution equation in physical space (\ref{eqn:scalar_general_evolution_equation}), we evaluate the spectral expansion of $\Theta(L_{\pm}^{A})$ in equation (\ref{eqn:L_pm_spectral_expansion}) at time $t$; this means we need to know $\mu_{\pm}$, $\nu_{\pm}$ and $\tau$ at time $t$. These can be recovered from the scattering functions $\boldsymbol{\psi}$ and $\boldsymbol{\phi}$ which are related to $\mathbf{N}$ and $\mathbf{M}$ by equation (\ref{eqn:jost_definition}). We first recover the Jost functions. The $\mathbf{N}$ function can be constructed from $\mathbf{N}_{1}$ using equation (\ref{eqn:N_inverse_scattering_integral_equation}) which gives
\begin{align}
    \mathbf{N}(x,k,t) = \begin{pmatrix}0\\1\end{pmatrix} + \frac{2 i \eta e^{- z_{t}(x)}}{k + i \eta} \sigma^{-1} \mathbf{N}_{1}(x,t) = \left( -i \eta\frac{\text{sech}\{z_{t}(x)\}}{k + i \eta},  \frac{k + i \eta \text{tanh}\{z_{t}(x)\}}{k + i \eta}\right)^{T}.
\end{align}
We can find the $\mathbf{M}$ Jost solutions in a similar manner to $\mathbf{N}$ using equations (\ref{eqn:M_inverse_scattering_integral_equation}) and (\ref{eqn:M_ell_inverse_scattering_integral_equation}) noting that $\tilde{C}_{1}(0) = 2 i \eta e^{-2 \eta x_{0}}$ and so $\tilde{C}_{1}(t) = 2 i \eta e^{-2 \eta x_{0} + 2 \eta \Theta(-\eta^2) t}$; this is $C_{1}(t)$ with $\eta$ replaced by $-\eta$. We find
\begin{align}
    M_{1}^{(1)}(x,t) = \frac{1}{2} \left(1 -  \text{tanh}\{z_{t}(x)\} \right), ~~
    M_{1}^{(2)}(x,t) = -\frac{1}{2} \text{sech}\{ z_{t}(x)\},
\end{align}
\begin{align}
    \mathbf{M}(x,k,t) = \left(\frac{k - i \eta \text{tanh}\{z_{t}(x)\}}{k + i \eta}, -i \eta\frac{\text{sech}\{z_{t}(x)\}}{k + i \eta}\right)^{T}.
\end{align}
From these, we can construct $\boldsymbol{\psi}$ and $\boldsymbol{\phi}$ using equation (\ref{eqn:jost_definition}). Using the wronskian relation in equation (\ref{eqn:AKNS_wronskian_relations_a}) and the fact that $\tau = 1/a$, we have
\begin{align}
    \tau(k) = \frac{k + i \eta}{k - i \eta}.
\end{align}
We then build $\mu_{-}$ and $\nu_{-}$, the eigenfunctions for the scalar system. These are
\begin{align}
    \mu_{-}(x,k,t) &= e^{2 i k x} \frac{\left( \eta^2 - k^2 + 2 i k \eta \text{tanh}\{z_{t}(x)\} \right)}{\left(k + i \eta\right)^2}, \label{eqn:one_soliton_eigenfunction_mu} \\
    \nu_{-}(x,k,t) &= e^{-2 i k x} \frac{\left( k^2 - \eta^2 - 2 i k \eta \text{tanh}\{z_{t}(x)\} + 2 \eta^2 \text{sech}^2\{z_{t}(x)\} \right)}{\left(k + i \eta\right)^2}. \label{eqn:one_soliton_eigenfunction_nu}
\end{align}
Therefore, we can construct the kernel $g_{-}(x,y,k,t) = \tau^2(k) \nu_{-}(x,k,t) \mu_{-}(y,k,t)/\pi$ inside of the spectral definition of $\Theta(L_{-}^{A})$ in equation (\ref{eqn:L_pm_spectral_expansion}). We can now show that the soliton solution for $q$ given in equation (\ref{eqn:one_soliton_solution}) is truly a solution to the general evolution equation (\ref{eqn:scalar_general_evolution_equation}) by explicitly demonstrating that equation (\ref{eqn:scalar_general_evolution_equation_expansion}) holds. We will evaluate the operator 
\begin{align}
    \Theta(L_{-}^{A}) \partial_{x} q(x,t) = \int_{\Gamma_{\infty}^{(1)}} dk \Theta(k^2) \int_{-\infty}^{\infty} dy \,  g_{-}(x,y,k,t) \partial_{y} q(y,t), \label{eqn:one_soliton_verification_relation}
\end{align}
and show that it is equivalent to $-q_{t}$ where $q$ is defined by equation (\ref{eqn:one_soliton_solution}). It is most prudent to split this into its continuous and discrete parts as in equation (\ref{eqn:L_pm_spectral_expansion_continuous_discrete_split}). As we will see, the portion of the operator related to the continuous spectra will vanish while that associated to the single eigenvalue $k_{1}$ will satisfy the above equation. This comes from the integral
\begin{align}
    I(k,t) = \int_{-\infty}^{\infty} \mu_{-}(y,k,t) \partial_{y} q(y,t) \, dy = 0, \label{eqn:vanishing_integral}
\end{align}
for all $k$. Looking at equation (\ref{eqn:L_pm_spectral_expansion_continuous_discrete_split}), we can see that this statement implies that the continuous portion, the integral of $g_{-}(x,y,k)$ over the real line, vanishes. It also implies that the integrals associated to the discrete kernels $g^{(2)}_{-,1}(x,y)$ and $g^{(3)}_{-,1}(x,y)$ and the first half of the $g^{(1)}_{-,1}(x,y)$ integral are zero. The single term that does not vanish is
\begin{align}
    \Theta(L_{-}^{A}) \partial_{x} q(x,t) = - \frac{2 i}{(a_{1}')^2} \Theta(-\eta^2) \nu_{-}(x,i \eta,t) \int_{-\infty}^{\infty} \partial_{k} \mu_{-}(y,k,t)|_{k=i \eta} \partial_{y} q(y) \, dy.
\end{align}
Hence, we expect $-q_{t}(x,t) = \Theta(L_{-}^{A}) \partial_{x} q(x,t)$. By making the change of variables $\xi = z_{t}(y)$, $d\xi = z_{t}'(y) dy = 2 \eta dy$ where $z_{t}(x) = 2 \eta (x - x_{0}) - 2 \eta \Theta(- \eta^2) t$, the above integral becomes
\begin{align}
    \int_{-\infty}^{\infty} \partial_{k} \mu_{-}(y,k,t)|_{k=i \eta} \partial_{y} q(y) \, dy = -i e^{z_{t}(x) - 2 \eta x} \int_{-\infty}^{\infty} \left(2 \eta x - z_{t}(x) + \xi\right) \text{sech}^2\xi \text{tanh}\xi \, d\xi.
\end{align}
Of the three terms in the integral, the first two vanish because the function is odd. The final term, i.e., $\xi \text{sech}^2\xi \text{tanh}\xi$, can be evaluated using integration by parts and the fundamental theorem of calculus to give
\begin{align}
    \int_{-\infty}^{\infty} \xi \text{sech}^2\xi \text{tanh}\xi \, d\xi = \frac{1}{2}\int_{-\infty}^{\infty} \text{sech}^2\xi \, d\xi = 1
\end{align}
Therefore, evaluating $\nu_{-}(x,k,t)$ in equation (\ref{eqn:one_soliton_eigenfunction_nu}) at $i \eta$ and using $(a'_1)^2 = - 4 \eta^2$ we find
\begin{align}
    \Theta(L_{-}^{A}) \partial_{x} q(x,t) = -4 \eta^2 \Theta(-\eta^2) \text{sech}\{z_{t}(x)\} \text{tanh}\{z_{t}(x)\}.
\end{align}
Comparing this to $q_{t}$
\begin{align}
    q_{t}(x,t) = 4 \eta^2 \Theta(-\eta^2) \text{sech}\{z_{t}(x)\} \text{tanh}\{z_{t}(x)\},
\end{align}
we notice that equation (\ref{eqn:one_soliton_verification_relation}) is true provided equation (\ref{eqn:vanishing_integral}) holds, which we will now confirm. Again, we make the change of variables $\xi = z_{t}(y)$ so that the integral $I(k,t)$ becomes
\begin{align}
    I(k,t) = A(k,t) \int_{-\infty}^{\infty} \tilde{\mu}_{-}(\xi,k) q'(\xi) \, d\xi
\end{align}
where $A(k,t) = e^{2 i k x_{0} + 2 i k \Theta(- \eta^2) t}$ and
\begin{align}
    \tilde{
    \mu}_{-}(\xi,k) &= e^{i k \xi/\eta} \frac{\left( k^2 - \eta^2 + 2 i k \eta \text{tanh}\xi \right)}{\left(k + i \eta\right)^2}, \\
    q'(\xi) &=  -2 \eta \, \text{sech}\xi \text{tanh}\xi.
\end{align}
If we introduce
\begin{align}
    I_{1}(k) &= \int_{-\infty}^{\infty} e^{i k \xi/\eta} \text{sech}\xi \text{tanh}\xi \, d\xi, \\
    I_{2}(k) &= \int_{-\infty}^{\infty} e^{i k \xi/\eta} \text{sech}\xi \text{tanh}^2\xi \, d\xi,
\end{align}
$I(k,t)$ can be written as
\begin{align}
    I(k,t) = -A(k,t) \frac{2 \eta}{(k + i \eta)^2} \left[ (k^2 - \eta^2) I_{1}(k) + 2 i k \eta I_{2}(k) \right].
\end{align}
Using integration by parts, we can put $I_{1}$ in terms of the Fourier transform of the hyperbolic secant.
\begin{align}
    I_{1}(k) = - \int_{-\infty}^{\infty} e^{i k \xi/\eta} \partial_{\xi} \text{sech}\xi \, d\xi = \frac{i k}{\eta} \int_{-\infty}^{\infty} e^{i k \xi/\eta} \text{sech}\xi \, d\xi
\end{align}
This Fourier transform can be evaluated by forming a rectangular contour in the complex plane with corners at $z = -R$, $z = R$, $z = R + i \pi$ and $z = - R + i \pi$. Integrating around this contour, the pole at $z = i \pi/2$ is the only residue. Then --- in the limit $R\to\infty$ --- the left and right sides of the contour integral vanish, and the top contour can be written in terms of the bottom contour (which is the integral along the real line we want when $R\to\infty$) because $e^{i k (\xi+i\pi)/\eta} \text{sech}(\xi+i\pi) = - e^{- k \pi/\eta} e^{i k \xi/\eta} \text{sech}\xi$. With these notes, $I_{1}$ is found to be
\begin{align}
    I_{1}(k) = i\frac{k \pi}{\eta} \text{sech}\frac{k \pi}{2 \eta}. \label{eqn:I1_integral_value}
\end{align}
Then, using a tricky manipulation, $I_{2}$ can be written in terms of $I_{1}$ as
\begin{align}
    I_{2}(k) = \left[\frac{i k}{2 \eta} + \frac{\eta}{2 i k}\right] I_{1}(k). \label{eqn:I2_integral_value}
\end{align}
Taking equations (\ref{eqn:I1_integral_value}) and (\ref{eqn:I2_integral_value}), we find that
\begin{align}
    I(k,t) = 0.
\end{align}
Therefore,
\begin{align}
    \Theta(L_{-}^{A}) \partial_{x} q(x,t) = - q_{t}(x,t),
\end{align}
and the one soliton in (\ref{eqn:one_soliton_solution}) is a solution to the general evolution equation in (\ref{eqn:scalar_general_evolution_equation}).

\subsection{The One Soliton for fmKdV and fSG}

The general nonlinear evolution equation becomes the fmKdV equation when we put $\Theta(L_{-}^{A}) = - 4 L_{-}^{A} |4 L_{-}^{A}|^{\epsilon}$ and it becomes the fsineG equations with $\Theta(L_{-}^{A}) = \frac{|4L_{-}^{A}|^{\epsilon}}{4 L_{-}^{A}}$ where $0 < \epsilon < 1$. Therefore, for these two equations, the one-soliton solution given in (\ref{eqn:one_soliton_solution}) becomes
\begin{align}
    q_{m}(x,t) &= 2 \eta \, \text{sech}\left\{2 \eta (x - x_{0}) - (2 \eta)^{3+2\epsilon} t\right\}, \\
    q_{SG}(x,t) &= 2 \eta \, \text{sech}\left\{2 \eta (x - x_{0}) + (2 \eta)^{-1 + 2 \epsilon} t\right\}.
\end{align}
We also find the "kink" solution $u$ from $q_{SG} = u_{x}/2$ to be
\begin{align}
    u(x,t) = \arctan{\text{sinh}\left\{2 \eta (x - x_{0}) + (2 \eta)^{2\epsilon-1} t\right\}}.
\end{align}
Both solutions are traveling waves which propagate without dissipating. The speed of the two solitons are given by
\begin{align}
    v_{m}(\eta) &= (2 \eta)^{2 + 2 \epsilon} \label{eqn:fmKdV_velocity} \\
    v_{SG}(\eta) &= (2 \eta)^{-2 + 2 \epsilon} \label{eqn:fsineG_velocity}
\end{align}
Notice that the fractional equations predict power law relationships between the speed of the wave and the amplitude of the wave, $\eta$. Therefore, the fmKdV and fsineG equations predict anomalous dispersion, showing that this is a common characteristic of fractional nonlinear systems.

\begin{figure}[H]
\begin{centering}
\includegraphics[width=0.6\textwidth]{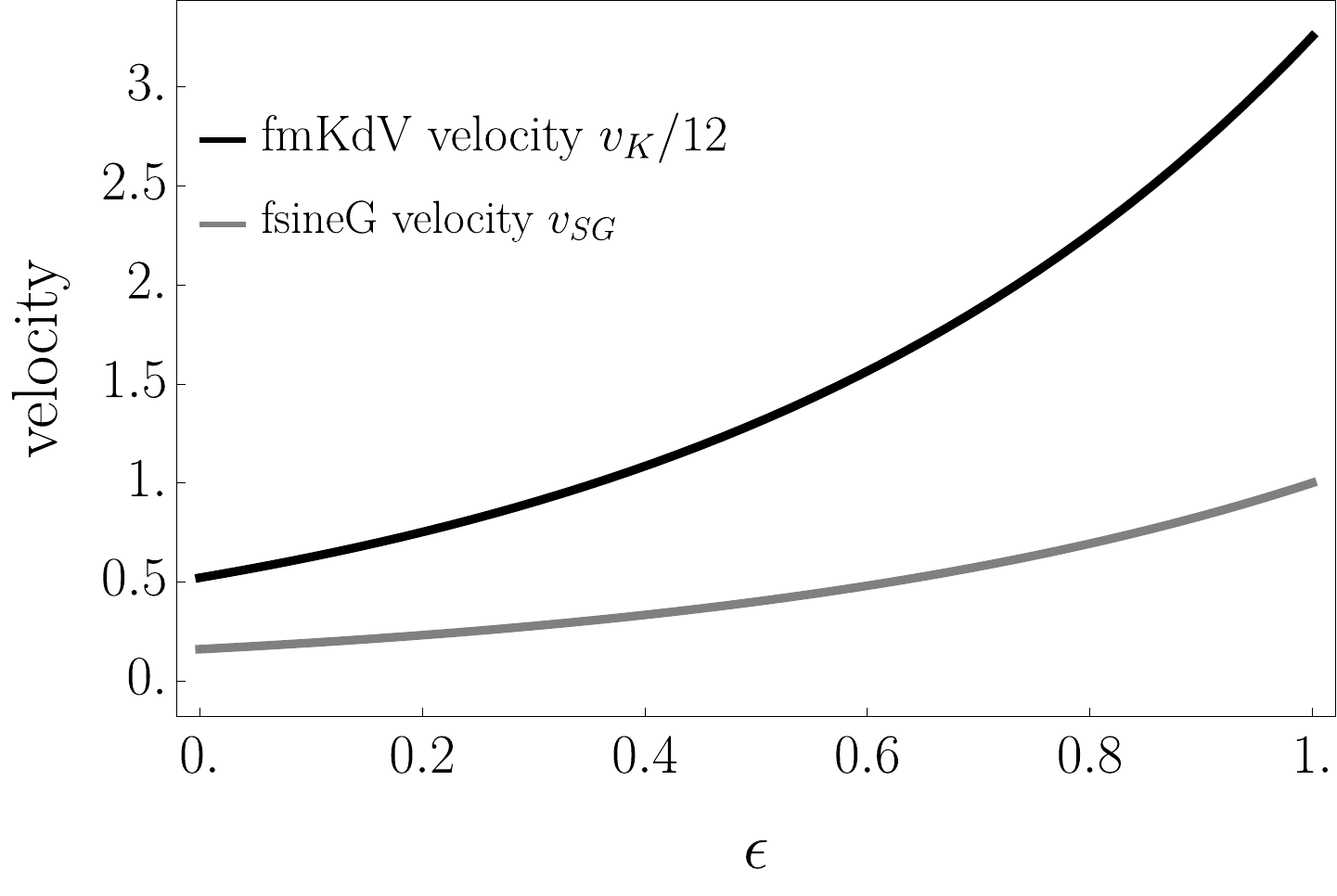}
\caption{\label{fig:velocity_plots}
    Localized waves predicted by the fmKdV and fsineG equations, equations (\ref{eqn:fmKdV_velocity}-\ref{eqn:fsineG_velocity}), show super-dispersive transport as their velocity increases for $0 < \epsilon < 1$. Notice that the fmKdV velocity is scaled by $1/12$. Also, at $\epsilon = 1$, the resulting nonlinear equations are described by integer operators. Just as in anomalous diffusion where the mean squared displacement is proportional to $t^{\alpha}$, the velocity in anomalous dispersion is proportional to $A^{\epsilon}$, where $A$ is the amplitude of the wave. Here $\eta = 3/2$.
}
\end{centering}
\end{figure}

\section{Conclusion}

We developed fractional extensions of the modified KdV, sine-Gordon, and sinh-Gordon equations on the line with decaying data. This process requires three key steps: a general evolution equation solvable by the inverse scattering transformation, completeness of squared eigenfunctions, and an anomalous dispersion relation. We demonstrated these three elements by developing a scalar general evolution equation using a symmetry reduction of the AKNS scattering system. Then, we found the fmKdV, fsineG, and fsinhG equations as a special case of this general evolution equation using the anomalous dispersion relations of the linear fmKdV, fsineG, and fsinhG equations, respectively. From scattering theory for the AKNS system, we found squared eigenfunctions and their associated operators for the scalar scattering problem. We then re-expressed completeness of the AKNS system in terms of these scalar squared eigenfunctions to give a spectral representation of the operator $\Theta(L_{\pm})$ in the general evolution equation. We developed the direct scattering, time evolution, and inverse scattering for the scalar scattering system and used these to derive the one-soliton solution for fmKdV and fsineG. We used the completeness relation to verify that these one-soliton solutions were truly solutions of fmKdV and fsineG. Finally, we showed that the one-soliton solutions of fmKdV and fsineG have power law relationships between the soliton's amplitude and velocity. This super-dispersive transport is an experimentally testable prediction of this theory.

\section*{Acknowledgements}
We thank U. Al-Khawaja, J. Lewis, and A. Gladkina for useful discussions. This project was partially supported by NSF under grants DMS-2005343 and DMR-2002980.

\printbibliography

\end{document}